\documentstyle[12pt]{article}
\newcommand{\be}{\begin{equation}}
\newcommand{\ee}{\end{equation}}
\newcommand{\s}{\section}
\newcommand{\ci}{\cite}
\newcommand{\r}{\ref}
\begin{document}
\begin{titlepage}
\begin{flushright}
FTUV/92-2
\end{flushright}
\begin{center}

\vskip 2cm
{\Large {\bf About the realization of chiral symmetry in $QCD_2  ^*$}}\\

\vskip 1.7cm
{\large {A. Ferrando$^{\dag}$}}

\vskip 0.2cm
{Institute for Theoretical Physics}\\
{University of Bern}\\
{Sidlerstrasse 5, CH-3012 Bern, Switzerland}\\
\vskip 0.2cm
and

\vskip 0.3cm

{\large {V.Vento $^{\ddag}$}}

\vskip 0.2cm
{Departament de F\'{\i}sica Te\`{o}rica and I.F.I.C.}\\
{Centre Mixt Universitat de Val\`{e}ncia -- C.S.I.C.}\\
{E-46100 Burjassot (Val\`{e}ncia), Spain.}

\vspace{1.8cm}
{\bf Abstract}
\vspace{0.2cm}
\begin{quotation}
{\small Two dimensional massless Quantum Chromodynamics presents
many features which resemble those of the true theory. In
particular the spectrum consists of mesons and baryons arranged
in flavor multiplets without parity doubling. We analyze the
implications of chiral symmetry, which is not spontaneously broken
in two dimensions, in the spectrum and in the quark condensate.
We study how parity doubling, an awaited consequence of Coleman's
theorem, is avoided due to the dimensionality of space-time and
confinement. We prove that a chiral phase transition is not possible
in the theory.}
\end{quotation}
\end{center}
\vspace{2.0cm}
$^*$Supported in part by CICYT grant \# AEN90-0040 and DGICYT
grant \# PB88-0064. \\
$^{\dag}$ Post Doctoral Fellow of CICYT- Plan Nacional de Altas
Energ\'{\i}as. \\
$^{\ddag}$ Vento @ Evalun11.bitnet
\end{titlepage}
\baselineskip  0.2in

\s{Introduction}
In recent work  we have obtained a description of $QCD_2$ in
terms of bosonic fields by establishing through the bosonization
rules a correspondence between characteristic Green functions of
the bosonized theory and well known fermionic amplitudes
\ci{fv91}. We have discovered the crucial role of the chiral
sector, which consists loosely speaking of free zero mass mesons,
in attributing quantum numbers. These so-called chiral mesons are
intimately related to the realization of chiral symmetry \`a la
Berezinkii-Kosterlitz-Thouless \ci{bk70} and are therefore no
Goldstone bosons. The excited mesons have their origin in the
gluonic interactions of the colored meson sector. In the chiral limit,
the dynamics of the two sectors is completely decoupled. The only
{\em true} baryons of the theory are also massless \ci{am82}.
Their genesis requires that the chiral sector be endowed with
topologically non-trivial boundary conditions. The remaining
baryons arise due to an interplay between the interactions in
the colored meson sector and the topological non-trivial boundary
conditions of the chiral sector. Therefore they appear as pseudo-mesonic
excitations on top of the {\em true} baryons.

The aim of this paper is to complete this description, discussing
those issues which have been a matter of debate over the years.
We start by recalling the properties of the spectrum and by
constructing explicitly the creation operator for the massless
baryons using abelian bosonization. It is a matter of notation to
establish a connection with our earlier work. We next review the
$BKT$ phenomenon and 't Hooft's consistency condition in order to
set the ground for our discussion on parity doublets. According
to Coleman's theorem \ci{co73} in two dimensions chiral symmetry
cannot be spontaneously broken and consequently one would expect
parity doublets in the spectrum, which do not arise \ci{ho74}.
We shall show that the dimensionality of space-time and
confinement are responsible for their absence.
Finally arguments based on the large number of color approximation
lead to a non-vanishing quark condensate and suggest, if naively
interpreted, a phase transition between the weak and strong
coupling regimes of the theory, purely on chiral symmetry
arguments. We interprete this result and confirm
Coleman's theorem for any $N$.

We shall hereafter consider only one flavor. This restriction is
irrelevant in the chiral limit of the theory, i.e., for massless
quarks. Under these circumstances the basis of our
argumentation, namely the separation between the flavor and the color
degrees of freedom remains. Therefore all our
discussion can be easily generalized to the many flavor case.

\s{The spectrum}
By bosonizing $QCD_2$ we have obtained an equivalent theory with
a massless meson \ci{fv91}. This so-called chiral meson is
completely decoupled from the remaining fields of the theory and
is not affected by the interactions. It is suggestive of a
Goldstone boson, which however cannot be, due to Coleman's
theorem. The remaining meson spectrum is described by means of
colored mesons, leading to a stable, discrete and infinite
spectrum with no parity doublets \ci{fv91,ho74}.

The interpretation of the baryonic spectrum in the bosonized
theory relies on the existence of zero mass baryons. In the
fermionic description they arise due to the kinematics of the two
dimensional world \ci{am82}. In our bosonic description they are
a consequence of the non-triviality of the boundary conditions of
the meson fields \ci{ef81,fv92}, in a way which we next describe.

In abelian bosonization \ci{ma75} the following solitonic
operators may be constructed for a non-abelian theory
\be
s_{\pm}(x,t) = : \exp{\left(-i \sqrt{\frac{\pi}{2}}\left(
\int^x_{-\infty} d\xi \dot{\varphi}(\xi,t) \pm
\varphi(x,t)\right)\right)}:
\ee
and
\be
s^{\alpha}_{\pm}(x,t) = : \exp{\left(-i \sqrt{\frac{\pi}{2}}
(-)^{\alpha}\left( \int^x_{-\infty} d\xi \dot{\eta}(\xi,t) \pm
\eta(x,t)\right)\right)}: \;\; , \;\; \alpha = 1,2
\ee
Here $\varphi$ represents the chiral meson field, a singlet under
the non-abelian group, which we shall take for simplicity to be
in this section $SU(2)_C$; $\eta$ is, loosely speaking, a color meson which
defines an extraordinary non-local representation of the color
group \footnote{We follow in this section the work of Halpern \ci{ha75} very
closely. One can look up in it the properties of the non-local representation
associated with the $\eta$ field. The boundary conditions on this field
play a very important role in the definition of its transformation
properties under the color group, see also \ci{fv92} for a
discussion on this point.} ; $\pm$ represent the corresponding chiralities
\ci{wi84}; $s^{\alpha}, \alpha = 1,2$ transforms as a doublet
under the color group. We shall use negative chirality fields
from now on unless specifically labelled.

If $\varphi$ and $\eta$, which we shall represent collectively by
$\phi$, are endowed with non-trivial boundary conditions
\be
\phi(\infty) =\sqrt{\frac{\pi}{2}} \;\; ; \;\;\;\;\;  \phi(-\infty) = 0
\ee
the $s$, $s^{\alpha}$ operators represent the $U(1)$ and
$SU(2)_C$ solitons respectively. Adscribing their winding numbers
properly normalized to baryon number ($B$) and color charge
($T_3$) respectively we have for these solitons and for the
quarks fields
\be
q^{\alpha}(x) = \sqrt{k\mu} s(x)s^{\alpha}(x) , \;\;\; \alpha
= 1,2
\ee
where $k$ is a numerical constant and $\mu$ a normal ordering
mass, the quantum numbers shown in Table 1
\begin{center}
\begin{tabular}{|c|c|c|c|c|c|} \hline
      & $s$ & $s^1$ & $s^2$ & $q^1$ & $q^2$ \\ \hline
  $B$ & $\frac{1}{2}$ & $\;0$ & $\;0$ & $+\frac{1}{2}$ & $+\frac{1}{2}$ \\
 \hline
  $T^3$ &  0 & $-\frac{1}{2}$ & $+\frac{1}{2}$ & $- \frac{1}{2}$ &
  $+ \frac{1}{2}$ \\ \hline
\end{tabular}
\end{center}
\vskip 0.2cm
\begin{center}
Table 1: Quantum numbers of solitons and quarks
\end{center}

It should be noted that despite the complicated transformation
properties of the $\eta$ field, the quark field transforms as
the fundamental representation of the color group. Abelian
bosonization leads to a highly non-trivial theory, where
the off-diagonal currents are spatially non-local.
However the baryon currents for zero mass quarks are free
\ci{ha76}.

Let us construct the analog of the four dimensional baryon
currents \ci{io81,et83}
\be
B(x) =\varepsilon^{\alpha \beta} q_{\alpha}(x) q_{\beta}(x)
\ee
Since $ [\varphi,\eta] = 0 $ its bosonized equivalent is given by
\be
B(x) = k\mu s(x)s(x)s^1(x)s^2(x)
\ee
Using  $:e^C::e^D: = e^{[C^+,D^-]}:e^{C+D}:$, where the
superindices $\pm$, as is custumary in field theory, indicate
the sign of the energy exponents in the field expansion, and the free
commutation relations for the boson fields \ci{ma75} we obtain
\be
B(x,t) \stackrel{\varepsilon \rightarrow 0}{\approx}
k\mu e^{[A^+(x+\varepsilon,t),A^-(x,t)]_{\varphi}}
e^{-[A^+(x+\varepsilon,t),A^-(x,t)]_{\eta}}:e^{2A[\varphi]}:
1_{\eta}
\label{baryon}
\ee
where
\be
A^{\pm}[\phi] = -i\sqrt{\frac{\pi}{2}}\left(\int^x_{-\infty}
d\xi \dot{\phi}^{\pm} (\xi,t) - \phi^{\pm}(x,t) \right)
\ee
Since for free fields the $\varphi$ and $\eta$ commutators are
identical we obtain
\be
B(x,t) = k\mu : \exp{\left(-i \sqrt{2\pi}
\left(\int^x_{-\infty} d\xi \dot{\varphi}(\xi,t) -
\varphi(x,t)\right)\right)}:
\ee
which certainly creates a baryon number one state.
The mass bosonization equation \ci{fv91} tells us,
both in the abelian and non abelian schemes,
\be
q^{\alpha+}_- q_{A+} = k \mu s^+_-s_+s^{\alpha +}_-s_{A+} = k \mu
e^{i\sqrt{2\pi}\varphi} g^{\alpha}_{A}
\label{mass}
\ee
Since
\be
s^+_-s_+ = e^{i\sqrt{2\pi}\varphi}
\ee
then
\be
s^{\alpha +}_- s_{A+} = g^{\alpha}_A
\ee
Thus an exact relation can be found between the
matrix elements of the non-abelian color fields of ref.(\ci{fv91})
and those of the soliton operator. Using the exact solutions of the
free and Wess-Zumino-Witten theory \ci{kz84} we obtain
\be
<e^{i\sqrt{2\pi}\varphi(x,t)} e^{-i\sqrt{2\pi}\varphi(0)}> =
<g^{1+}_1(x,t) g^1_1(0)> = \frac{1}{2\pi k\mu \sqrt{x_+x_-}}
\ee
{}From these expressions one can calculate the
correlators for the solitonic operators which are given by
\be
<s^+_{\pm} (x,t) s_{\pm} (0)> = <s^{\alpha+}_\pm
(x,t) s^{\alpha}_{\pm} (0)> = \left( \frac{-i}{2\pi k \mu
x_{\mp}} \right)^{\frac{1}{2}}
\label{scorrelator}
\ee
leading to the baryon correlator
\be
<B^+(x,t) B(0)> = \frac{-1}{4\pi^2 x^2_+}
\ee
which corresponds to a zero mass particle. $B(x,t)$ projects onto
a zero mass baryon number one state.

We have throughout assumed free $\varphi$ and $\eta$ fields. This
is consistent with the fact that the interaction is
asymptotically free \ci{ha76}. Looking back at
the derivation we see in Eq.(\r{baryon}), that because the fields
are free the color dependence banishes, establishing that the
baryon number one soliton current associated with the massless
baryon, depends solely on the chiral field.

Let us recover our previous result \ci{fv91} in this more transparent
language. The baryon correlation function is terms of the
$\varphi$ and $\eta$ meson fields is given by
\be
G^B(x_a,x_b;y_a,y_b) = k^2\mu^2 <B^{\varphi +}(x_a,x_b)
B^{\varphi}(y_a,y_b)><B^{\eta +}(x_a,x_b)
B^{\eta}(y_a,y_b)>
\ee
If one uses a specific gauge \ci{ei76} such that
\be
B^{\varphi} = s(x_a) s(x_b) \;\;\;\; and  \;\;\;\; B^{\eta} =
\varepsilon_{\alpha \beta} s^{\alpha}(x_a) s^{\beta}(x_b)
\ee
it is easy to show, that in the long distance regime
($l_a = x_a - y_a ; l_1 \rightarrow
\infty$) the zero mass particle dominates the $\varphi$ piece of
the correlator leading
to
\be
<B^{\varphi +}(x_a x_b) B^{\varphi}(y_a,y_b)> \sim
\frac{1}{l_1^2}
\ee
Since $G^B$ is the true propagator, its long distance behavior is
a sum of Yukawas. This together with the last equation implies
that the $B^{\eta}$ correlator cannot have a physical long
distance behavior and the intermediate states that saturate it
must be unphysical mesons \ci{fv91}.

\s{Coleman's theorem and parity doubling}
It is well known that in two dimensions chiral symmetry cannot be
spontaneously broken \ci{co73}. In $QCD_2$ it is realized however
not in a conventional way but in an {\em almost} spontaneously
broken fashion \ci{wi78}. This so-called BKT phenomenon is charaterized
by a correlation function that falls of as a power law \ci{fv91}
\be
<q \bar{q}(x) q \bar{q}(0)> \stackrel {x^2
\rightarrow \infty}{\approx} <e^{i\sqrt{\frac{4\pi}{N}}\varphi(x)}
e^{-i\sqrt{\frac{4\pi}{N}}\varphi(0)}> \approx \frac{1}{(2 \pi k^2
\mu^2 x^2)^{\frac{1}{N}}}
\label{qcorrelator}
\ee
where $N$ is the number of colors. Thus the chiral meson, although
not a Goldstone boson, governs the long distance behavior of this
correlator, a property shared with its four dimensional analog.
This similarity of behavior goes even further. G. 't Hooft
analyzed the restrictions imposed by the axial anomaly equation
on the spectrum of confining theories with zero mass quarks
\ci{ho79}. In two dimensions the anomaly appears by considering
the vector current (axial current) two point function (recall
$j^{\mu}_5 = \varepsilon^{\mu \nu}j_{\nu}$). The discontinuity of
this amplitude, a signature of the anomaly \ci{fs81}, can also be
calculated by saturating over all possible intermediate states.
In the chiral limit, only massless boundstates contribute at zero
momentum. In $QCD_2$ we have both zero mass mesons and baryons.
The massless mesons saturate the anomaly as occurs in the
spontaneousy broken theories \ci{ho79,bl82}. The zero mass baryons
do not contribute \ci{am82}.

Since chiral symmetry is not spontaneously broken the spectrum
should contain parity doublets. However they do not appear, again
in direct correspondence with the four dimensional situation.
However the mechanism which elliminates the doubling
must be very peculiar of two dimensions.

In the bosonized form of the theory the chiral and
color sectors are completely decoupled. The chiral sector
is generated by the zero mass meson field. Gauge interactions
are absent. Since chiral symmetry
is not broken one should expect, besides the $\varphi$ meson, a
chiral partner, i.e., a $\sigma$. The conservation of the
$U(1)_{axial}$ current leads to
\be
\partial_{\mu} j^{\mu}_{5} = 0
\label{axial}
\ee
while the dimensionality of space-time implies
\be
j^{\mu}_{5} = \varepsilon^{\mu \nu} j _{\nu}
\ee
Since $j^{\mu}_{5} \sim \partial ^{\mu} \varphi$, Eq.(\r{axial})
leads to the existence of a pseudoscalar zero mass boson
\be
\Box \varphi = \partial_{\mu} j^{\mu}_{5} = 0
\ee
Moreover the current associated with $U(1)_{vector}$
\be
j_{\mu} \sim \varepsilon_{\mu \nu} \partial ^{\nu} \varphi
\label{vector}
\ee
is trivially conserved and therefore no $\sigma$ meson is
required. Eq.(\r{vector}) is not a Noether current. It is
conserved irrespective of the dynamics and has therefore a
topological character. Thus no zero mass doublets exist!

The color sector is determined by the minimally coupled
Wess-Zumino-Witten model, which realizes by
construction $SU(N)_{vector}$ locally and therefore also
globally. The vacuum of the theory is also invariant under this
group since the symmetry cannot be broken spontaneously. On
the other hand $SU(N)\otimes SU(N)$ is explicitly broken. The breaking
is caused by the color anomaly, which in the bosonized language
appears explicitly in the action \ci{fv91,ad83}
\be
{\it S}_{WZW} (g,A_{\mu}) = {\it S}_{WZW} (g) + \int d^2x
tr(A_+J_- +A_-J_+ +\frac{1}{4} A_+A_- - \frac{1}{4\pi}
A_+gA_-g^+)
\ee
where ${\it S}_{WZW}$ is the $WZW$ action and
\be
J^a_{\mp} = -\frac{i}{2\pi} tr (\partial _{\mp} g^{\mp 1}) g^{\pm
1} T^a
\ee
The vacuum of the theory is therefore not invariant under
$SU(N)\otimes SU(N)$. Finally parity is a good symmetry of the
WZW model, because this is exactly what happens in $QCD$ \ci{wittenp}.

The spectrum of the resonant mesons arises from this action.
The symmetries of the hamiltonian and the properties of the vacuum
impose severe restrictions on the possible allowed quantum numbers and
degeneracies of these states. In two dimensions the representations of
Poincare's group do not require angular momentum and therefore only a
principal quantum number {\em n} associated with the energy of the
boundstate will characterize the space-time component of the wave
function. Parity, {\em P}, will also be a good quantum number of the states.

In two ($1+1$) dimensions degenerate opposite parity states can
only arise due to internal degrees of freedom. In massless
$QCD_2$ the internal degrees of freedom are flavor and color.
In the bosonized theory flavor only appears in the chiral sector,
which on the other hand is color blind. As we have seen, the
chiral sector avoids parity doubling through the relation between
the axial and vector currents, which arises as a consequence of
the dimensionality of space-time. Color describes the dynamics of
the WZW model and therefore of the other sector of the bosonized
theory. The fields describing this action are flavor singlets, and
therefore this theory is flavorless (insipid). A degeneracy
between states of different parity can only occur if the color
structure of the model allows it. This is not the case due to
confinement. Let us be more precise. Imagine for a moment that
$SU(N) \otimes SU(N)$ were not broken. In this scenario we would
have many degenerate multiplets of opposite parity. For example
in the case of two colors, the only mesonic representations which
satisfy confinement are of the form
\be
\left( \frac{m}{2}\;,\;\frac{m}{2} \right) \;\;,\; m \in {\bf\it N}
\ee
Their decomposition in terms of the vector color symmetry is
given by
\be
\left( \frac{m}{2}\;,\;\frac{m}{2} \right) \Rightarrow 0 \oplus
1\oplus 2 \oplus \cdots \oplus m
\ee
This multiplets contain in general states of different parity.
Recall for example the $(\sigma,\vec{\pi})$ representation of
$SU(2) \otimes SU(2)$. However there is only one singlet in each
representation. So, once we impose confinement, the degeneracy
will disappear. The argument generalizes to N colors.
\footnote{A caveat: $SU(N)\otimes SU(N)$ is explicitly
broken in the confining theory and one could argue, that the
breaking mechanism could make  opposite parity
singlets coalesce to the same multiplet. This can not be,
since it would imply, that  the symmetry is broken to a larger
group than  $SU(N)_{vector}$.}

Thus the dimensionality of space-time and the
realization of the color symmetry in the Wess-Zumino-Witten lagrangian
conspire to eliminate parity doubling, explaining
why the boundstates of the meson equation depend exclusively of
{\em n}, which also determines the parity {\em P} = $(-)^{n+1}$ \ci{ho74}.
Similar arguments can be developed for baryons.

\section{The chiral phase transition}

As we have seen in the previous section, Eq.(\r{qcorrelator}),
in the large $N$ limit the quark condensate does not vanish.
This has been considered by many people as a signature of
spontaneously broken chiral symmetry and therefore as a loophole
of Coleman's theorem \ci{zh85,cf88}. We next show that this
interpretation is incorrect.\footnote{A beautiful discussion
about the long distance behaviour of the chiral condensate
and its relation to the
$\frac{1}{N}$ expansion for the $SU(N)$ Thirring model
can be found in ref.\ci{wi78}.}

The cornerstone of the suspicion of a chiral phase transition is the
singular behavior of the quark condensate $<q\bar{q}>$, since
this expectation value is an order parameter for the breaking of
axial flavor symmetry. Taking the behavior of the condensate in
the large $N$ limit one can envisage a phase change from a
situation in which axial symmetry is conserved ($<q\bar{q}> = 0$),
to one in which it is broken ($<q\bar{q}> \neq 0$). Since 't
Hooft's calculation \ci{ho74} was performed in this latter case,
it could seem that his results would correspond to one of the
phases of the theory and therefore not extendeble to the other
regime. The authors which have taken this position consider 't
Hooft's calculation as the weak coupling limit of the theory,
since  the transition region occurs for $e'\sim M \rightarrow 0$, $M$
being the quark mass and
$e'$ is the effective coupling, $e'= \frac{e}{\sqrt{N}}$. If
one accepts this argumentation the physics for finite $N$ must be
substantially different from that of 't Hooft's model and
corresponds to a phase where chiral symmetry is restored
\ci{pa77}. Note that this way of thinking requires an
$N$ dependence of the quark mass, which we consider
academic, since quark masses are external parameters in
$QCD$ and moreover are associated with flavor, not with color.

Our study of bosonized $QCD_2$ allows us to clarify the above
statements. The exact calculation of the quark correlator shows
an anomalous behavior only if $M=0, N \approx \infty$. This
behavior is controlled (see Eq.(\r{qcorrelator})) by the chiral
sector of the theory. Let us repeat the argumentation, leading to
this formula for the chiral condensate. Using the bosonization
formula Eq.(\r{mass}) we obtain for it
\be
<q \bar{q}> \sim
<0|:e^{i\sqrt{\frac{4\pi}{N}}\varphi}:|0><0|tr(g)|0>
\label{condensate}
\ee
In this expression the vacuum expectation value (vev) of $tr(g)$ is
always different from zero, since the $SU(N) \otimes SU(N)$ symmetry is
explicitly broken. Thus only the vev of the chiral operator
$e^{i\sqrt{\frac{4\pi}{N}}\varphi}$ can be the cause of the
vanishing of the condensate. This is precisely what happens. The
chiral operator changes under transformations of the
$U(1) \otimes U(1)$ group, defined by $\varphi \rightarrow \varphi +
\alpha$ as
\be
e^{i\sqrt{\frac{4\pi}{N}}\varphi} \rightarrow
e^{i\sqrt{\frac{4\pi}{N}}(\varphi+ \alpha)}
\ee
If the vacuum is invariant under $U(1) \otimes U(1)$ (Coleman's
theorem) the vev must vanish. This vev is therefore the order
parameter in the bosonized theory for any phase transition
associated with chiral symmetry.

In the large $N$ limit the vev of the quark condensate does not
vanish. Naively this would imply that our order parameter does
not vanish and that the vacuum would break spontaneously the
symmetry. This argument is incorrect. In the large $N$ limit the
fluctuations of the $\varphi$ field disappear and
$e^{i\sqrt{\frac{4\pi}{N}}\varphi}$ becomes the unit operator, so
that the quark condensate becomes in this limit purely an $SU(N)$
object ($<tr(g)>$), which is different from zero due to the gauge
interaction. Thus our order parameter, which was such because it
transformed non trivially under the group, ceases to be an order
parameter in the large $N$ limit because it becomes the unit
operator which is trivially invariant under the group. In this
way, the fact that the condensate becomes different from zero in
the large $N$ limit is independent of chiral dynamics and simply
reflects the fact that $SU(N) \otimes SU(N)$ is explicitly broken.

\s{Conclusion}
We have used abelian and non-abelian bosonization techniques to
study the properties of two-dimensional $QCD$. The bosonization
rules have provided a well defined connection between the Green
functions of the bosonized models and those of their fermionic
counterparts. The crucial importance of the apparently naive
chiral sector has been unveiled, both for mesons and baryons, and
the role of the gauge dynamics in the formation of the spectrum
has been clarified.

The zero mass baryon is essential to describe the excited
baryons. Its existence is guaranteed by the presence of a
topological chiral sector. We have constructed its current
operator and described the properties of the baryonic spectrum
as obtained from the four point correlators and its absence
from the chiral anomaly.

The chiral meson is instrumental in describing the flavor
symmetry \`a la BKT. This mechanism converts the impossibility of
manifesting itself as the Goldstone boson of a spontaneously
broken symmetry, into a dominance of the long range behavior
of the correlators.

The resonant mesonic spectrum has its origin in the dynamics of
color. The non-existence of parity multiplets is due to the
dimensionality of space-time and confinement. The latter mechanism
is a consequence of the gauge coupling, which also leads to an
explicit breaking of the axial color symmetry in such a way as to
preserve parity and $SU(N)_{vector}$. In bosonized $QCD_2$ this
breaking appears as an
explicit term in the lagrangian, while in its fermionic
equivalent version it is due to the anomaly, and therefore a
consequence of the non-invariance of the fermonic measure under
the axial color symmetry.

We have analyzed the possibility of a phase transition in
$QCD_2$. The cornerstone of this proposal is the behavior of the
chiral condensate in the large $N$ limit. We have proven that
only one phase exists and that the apparent anomalous behavior of
the chiral order parameter is a consequence of the trivialization
of the chiral sector in that limit. We have shown that the
physics becomes independent of the chiral dynamics. We therefore
must conclude that all argumentation aiming at justifying the
existence of a phase transition in $QCD_2$, based on the spontaneous
symmetry breaking of the $U(1) \otimes U(1)$ symmetry is spurious and
interprets incorrectly the mechanisms regarding the symmetries of
$QCD_2$.

\s*{Acknowledgement}
We have received useful comments and criticism from M. Asorey, A.
Gonz\'alez-Arroyo, D. Espriu, J. Ros and A. Santamar\'{\i}a. One
of us (A.F.) is grateful to H. Leutwyler, P. Minkowski, A.
Smilga and U.J. Wiese for illuminating discussions.

\end{document}